\documentstyle [a4,11pt,amssymbols]{article}
\title{ {\bf The 1-Point Cluster Distribution Function and its Moments} }
\author{ {\bf Manolis PLIONIS}$^{1,2}$ \& {\bf Riccardo VALDARNINI}$^{2}$ \\
{\it $^1$ICTP - International Centre for Theoretical Physics,} \\
{\it Strada Costiera 11, 34013 Trieste, Italy} \\
{\it $^2$SISSA - International School for Advanced Studies,} \\
{\it Via Beirut 2-4, 34013 Trieste, Italy} }
\date{}

\begin{document}

\maketitle

\thispagestyle{empty}

\clearpage

\setcounter{page}{1}

\section*{\center Abstract}
We derive the 1-point probability density function of the smoothed
Abell-ACO cluster density field and we compare it with that of artificial
cluster samples, generated as high peaks of a Gaussian field in such a way
that they reproduce the low-order (2- and 3-point) correlation functions
and the observed cluster selection functions. We find that both real and
simulated {\em pdf}'s are well approximated by a log-normal distribution
even when the Gaussian smoothing radius is as large as 40 $h^{-1}$ Mpc.
Furthermore the low-order moments of the {\em pdf} are found to obey a
relation $\gamma \propto \sigma^4$, with $\gamma$ being the skewness. Since
clusters have not had enough time to significantly depart from their
original birth-place positions, these results are consistent with them
being high-peaks of an underlying initial Gaussian density field.

A by-product of our analysis is that when we rescale the {\em pdf} cluster
moments to those of the QDOT-IRAS galaxies, using linear biasing with
$b_{cI}\sim 4.5$ and for the common smoothing radius of 20 $h^{-1}$ Mpc, we
find them to be significantly smaller than those directly estimated from the
QDOT data by Saunders et al. (1991).

\clearpage

\section{Introduction}
The study of the large-scale galaxy distribution is of fundamental
importance in modern cosmology. A complete knowledge of the present cosmic
density fluctuation field $\delta_m  \equiv
\frac{\rho - \langle\rho\rangle}{\langle\rho\rangle} $,
in the linear regime, would yield informations
on the shape of the spectrum of primordial perturbations which gave rise,
in the gravitational instability picture, to the present cosmic structures
and their distribution.
Within the framework of the most common galaxy formation theories, the
standard two assumptions are $(i)$ that the initial $\delta_m$ field is
described by a Gaussian distribution and $(ii)$ that some sort of biasing
mechanism is at work.
In the most popular biasing models, galaxies form at the high peaks of the
underlying matter field (Kaiser 1984; Bardeen et al. 1986). This model was
introduced to reconcile, among other things, the observed low values of
$\Omega$ with the inflationary model (Olive 1990).

Many statistical approaches have been used to study the distribution of
extragalactic structures. The most common one is the two--point
correlation function, $\xi(r)$, (cf. Peebles 1980) which has been found to
have the form:
\begin{equation}\label{eq:obs}
\xi(r) = (r/r_{\circ})^{-\gamma} ,
\end{equation}
with $r_{\circ}\approx 5$ $h^{-1}$ Mpc and $\gamma \approx 1.8$ from 10
$h^{-1}$ Kpc up to 10 $h^{-1}$ Mpc (Davis \& Peebles 1983). Cluster of
galaxies exhibit similar kind of correlations up to $\sim 50$ $h^{-1}$ Mpc
with the same slope and $r_{\circ}\approx 20 - 25$ $h^{-1}$ Mpc (cf.
Bahcall \& Soneira 1983; Klypin \&  Kopylov 1983). Another statistical
measure is the 3--point correlation function, for which a non-zero signal
implies non--Gaussian statistics. The analysis of both 2-D and 3-D samples
has shown that galaxies do have a non-zero connected 3-point correlation
function with a hierarchical coefficient, $Q\simeq 1$ (Groth \& Peebles
1977; Sharp, Bonometto \& Lucchin 1984). Similar results hold also for
clusters of galaxies (T\'oth et al. 1989, Jing \& Zhang 1989, Jing \&
Valdarnini 1991 [hereafter JV]; Borgani, Jing \& Plionis 1992).

These results are of great interest, since non-Gaussianity might arise either
from  gravitational clustering, or from biasing mechanisms, or
finally from intrinsic non-Gaussian statistics of the initial matter field.
A complete knowledge of the N--point correlation function would allow one,
using the linear biasing assumption, to recover the statistical properties of
 the $\delta_m$ field. For $N > 3$, however, this approach becomes impractical
due to the limited extension of the available extragalactic samples.

An alternative method, which is becoming increasingly popular, is the study
 of the probability density function ({\em pdf}).
Theoretically the knowledge of the N--point correlation function is
 equivalent to that of the {\em pdf} and its moments, but for real data
sets the latter are easier to measure.
Because of the discrete nature of the available samples of extragalactic
objects, and in order to obtain a continuous density field one is forced
to smooth the discrete distribution with some function (usually a top-hat
or a Gaussian one is used).
For a normally
distributed primordial density field and as long as the variance
$\sigma^2 \equiv \langle\delta^{2}\rangle$ is small, under the action of
gravity, the deviation of the initial {\em pdf} from the Gaussian shape is well
approximated by the Edgeworth expansion (Colombi 1994).

In the non--linear regime N-body simulations have shown the the {\em pdf}
becomes non-Gaussian as the clustering evolves
(Bouchet, Schaffer \& Davis 1991; Bouchet \& Hernquist
1992; Lahav et al. 1993; Juszkiewicz et al. 1993; Kofman et al. 1994).
For $\sigma \ll 1 $ the shape of the {\em pdf } is well fitted by a lognormal
distribution. The {\em pdf} of different galaxy samples has been estimated
by Bouchet et al. (1993), Gatza\~naga and Yokoyama (1993).
Kofman et al. (1994) have compared the {\em pdf} derived form CDM N-body
simulations with that of the IRAS sample and the one
recovered using the POTENT procedure with $\Omega=1$.
Their main conclusion is that, if galaxies trace the mass, the observed
{\em pdf} can be fitted using Gaussian initial conditions.

Furthermore, the time evolution of the moments of the {\em pdf} has been
studied both analytically and with N-body simulations (Goroff et al.
1986;  Coles \& Frenk 1991; Juszkiewicz et al. 1993; Juszkiewicz, Bouchet
\& Colombi 1993; Kofman et al. 1994; Colombi 1994). For a given filter and
smoothing radius, the normalized skewness of the matter distribution,
$S_3 \equiv \langle\delta^3\rangle/ \langle\delta^2\rangle^2 $, will depend
on the shape of the power spectrum $P(k)$. Juszkiewicz ,Bouchet \& Colombi
(1993), using
second order theory, found $S_3 \simeq 34/7 -(3+{\rm n})$ for a top-hat
filter and scale-free spectra. These results agree with those obtained from
N-body simulations (Kofman et al. 1994). The proportionality
$\langle\delta^3\rangle \propto \sigma^4$ holds also at a fixed time when
one changes the smoothing radius (Coles \& Frenk 1991; Kofman et al. 1994).

In this paper we estimate the {\em pdf} for the distribution of Abell+ACO
clusters of galaxies (Abell 1958; Abell, Corwin \& Olowin 1989). Since
clusters of galaxies are the brightest objects in the sky  they can probe
the very large scales, reliably up to $ \approx 300$ $h^{-1}$ Mpc. On these
scales the density fluctuations are well within the linear regime and the
deviations of the {\em pdf} from a Gaussian distribution is determined
either by the fact that they are biased tracers of the underlying field or
due to non-Gaussian initial conditions. The main aim of this paper is to
establish whether the present cluster {\em pdf} can be derived entirely
under the assumption that clusters form at the high peaks of a Gaussian
background field. The basic procedure is to compare the observed cluster
{\em pdf}  with that measured from the distribution of simulated clusters,
generated as peaks of a initially Gaussian background field and  having the
same boundaries and selection effects as the real data.

\section{Data Samples}
\subsection{Real cluster samples}

We apply our analysis to the combined Abell-ACO $R\ge 0$
cluster sample, as defined in Plionis \& Valdarnini (1991) [hereafter PV91]
and analysed in Plionis, Valdarnini \& Jing (1992) [hereafter PVJ]. The
northern sample, with dec$\ge -17^{\circ}$ (Abell), is defined by those
clusters that have
measured $z\lesssim 0.1$, while the southern sample, (ACO) with dec$\le
-17^{\circ}$, is defined by those clusters with $m_{10} \le 16.4$ (note that
with this definition and due to the availability of many new cluster redshifts
only 7 ACO clusters have $m_{10}$ estimated redshifts
from the $m_{10}-z$ relation derived in PV91). Both samples are limited in
Galactic latitude by $|b|\ge 30^{\circ}$.
The redshifts we have used are mainly from Struble \& Rood (1987) and
Postman, Huchra \& Geller (1992) with additions from taken Rhee \& Katgert
(1988) and Batuski et al. (1991), the original ACO paper, Vettolani et al.
(1989); Cappi et al. (1991) and Muriel et al. (1991) as well as by our own
cross-correlation
analysis of the Fairall \& Jones (1991) galaxy redshift catalogue with
the cluster catalogues. The total number of clusters
in our samples is 357 Abell and 157 ACO ones.

We take into account the effect of Galactic absorption, which is assumed
to follow the usual cosecant law:
\begin{equation}\label{eq:obs1}
P(|b|) = \mbox{dex} \; \left[ \alpha \left(1 - \csc |b| \right) \right]
\end{equation}
with $\alpha \approx 0.3$ for the Abell sample (Bahcall \& Soneira 1983;
Postman et al. 1989) and $\alpha \approx 0.2$ for the ACO sample
(Batuski et al. 1989), by weighting each cluster by $1/P(|b|)$, a correction
which explicitly assumes that the unobserved (obscured) clusters are correlated
with the observed ones, a reasonable assumption under the well known
correlation properties of clusters.

We have derived the cluster-redshift selection function, $P(z)$, by fitting
the cluster density, as a function of $z$ (cf. Postman et al. 1989; PVJ),
by:
\begin{equation}\label{eq:red}
P(z) \left\{ \begin{array}{ll}

  = 1  & \;\;\;\;\; \mbox{ for} \; \; z \le z_c \nonumber \\
\propto \exp(-z/z_{\circ}) & \;\;\;\;\; \mbox{for} \;\; z>z_{c}
	\end{array}
	\right.
\end{equation}
where $z_{c}$ is the maximum redshift at which the sample exhibits a roughly
constant density (for more details see PVJ).
We find $z_c \approx 0.079$ and 0.066 for the Abell and ACO samples, which
correspond to $R\sim 230$ and 190 $h^{-1}$ Mpc respectively.
Cluster comoving distances are estimated using the standard relation
(Mattig 1958):
\begin{equation}\label{eq:comv}
R = \frac{c}{H_{\circ} q_{\circ}^{2} (1 + z)} \left(q_{\circ}z + (1-q_{\circ})
(1-\sqrt{2 q_{\circ} z +1}) \right)
\end{equation}
with $H_{\circ} = 100 \; h \;$ km sec$^{-1}$ Mpc$^{-1}$ and $q_{\circ} = 0.2$.

The Abell and ACO cluster densities, out to their limit of completeness
($z_c$), is $\sim 1.4 \times 10^{-5}$ $h^{3}$ Mpc$^{-3}$ and
$\sim 2.1 \times 10^{-5}$ $h^{3}$  Mpc$^{-3}$, corresponding to a
characteristic length scale of $\langle \rho \rangle^{-\frac{1}{3}} \approx
41$ $h^{-1}$ Mpc and 36 $h^{-1}$. The higher space-density of ACO clusters is
partly due to the huge Shapley concentration (Shapley 1930; Scaramella et
al. 1989; Raychaudhury 1989), but a significant part is also due to systematic
density differences between the Abell and ACO cluster samples, as a function
also of $z$, which have been noted in  a number of studies
(cf. PV91 and references therein) and which could be attributed to the high
sensitivity of the IIIa-J emulsion plates.

Since we want to use a cluster sample covering the whole sky we need to take
into account the density variations of the Abell and ACO cluster samples. We
will do this by renormalizing the ACO density to the Abell one using a radial
{\it matching} function, $W(R)$, as in PV91. Since in this paper we are mostly
interested in the cluster density fluctuations $\delta \rho/\rho$, it is
equivalent to
normalize the density of the Abell to the ACO or vice-versa. In order to test
the robustness of the results in the choice of the $W(R)$ function, we will
use a number of such functions, among which the $W(R)=1$ (no matching) and
the $W(R)=constant$ (no radial dependance) cases.

Note that in PVJ we had estimated the 2-point correlation function
for our samples of clusters and we found that the slope of the derived
2-p function has a value $\sim 1.8 \pm 0.2$
for both Abell and ACO samples. However their amplitudes
are slightly different, with Abell clusters having $r_{\circ} \simeq 18 \pm 4$
$h^{-1}$ Mpc (bootstrap errors used) out to $\lesssim 50$ $h^{-1}$ Mpc while
the ACO clusters have $r_{\circ} \simeq 22 \pm 10$ $h^{-1}$ Mpc but only out to
$\sim 30$ $h^{-1}$ Mpc. This could be, however, due to the fact that the ACO
sample is defined in a relatively small solid angle and therefore the large
wavelengths may be undersampled.
Based on our previous analysis (Jing et al. 1992), we believe that the
cluster correlations are reliable and mostly unaffected by cluster
contamination effects ({\em cf.} Sutherland 1988).

In PVJ we also estimated the 3-point correlation function using the moment
method (cf. Sharp, Bonometto \& Lucchin 1984; Jing \& Valdarnini 1991;
Borgani, Jing \& Plionis 1991). We found a good fit to the spatial 3-point
function provided by the hierarchical expression:
\begin{equation}
\zeta (r_1,r_2,r_{12})~=~Q\,[\xi(r_1)\xi(r_2)+\xi(r_1)\xi(r_{12})+
\xi(r_2)\xi(r_{12})]\,,
\label{eq:zetas}
\end{equation}
with $Q_{Abell} \approx 0.6$ and $Q_{ACO} \approx 0.7$,
with relative bootstrap error of $\sim \pm 0.2$.

Finally, in order to
minimize the uncertainties due to the approximate character of the redshift
selection function, $P(z)$, and of the radial function, $W(R)$, especially
at large distances, we will restrict our analysis to $R_{max} = 240$
$h^{-1}$ Mpc.

\subsection{Simulated cluster samples}

The procedure used to generate our artificial cluster catalogues has been
presented in PVJ. Here we only remind the reader the main steps of
our procedure.

We generate our simulated cluster catalogues using a method similar to
that of Postman et al. (1989). We identify the position $\vec{x}$
of points corresponding to clusters of galaxies according to the following
prescription:
$N_p$ points are randomly placed in a cube of size 640 $h^{-1}$ Mpc
with $N_g=64^3$ grid points. We keep points only within the sphere of radius
$r_{max} = 320$ $h^{-1}$ Mpc centered at the cube centre.
The points are then displaced from their original positions $\vec{x}_r$
using appropriately the Zel'dovich approximation.
The power spectrum of the density fluctuations, $P(k)$, has a Gaussian
distribution with random phases. $P(k)$ is chosen so that the perturbed
particle positions give a correlation function in the desired range. In
accordance with Postman et al. (1989), we use the following power spectrum:
\begin{equation}\label{eq:sim}
P(k)=A \; k^{\rm n} \; \exp(-|\vec k|^2/\Lambda^2) \; \Theta(|\vec k|)
\end{equation}
where $\Lambda^{-1}=0.1$ $h^{-1}$ Mpc, $\Theta(|\vec k|)=0 $ for
 $|\vec k| > 2\pi/80$ $h^{-1}$ Mpc, $A$ is a normalization constant and
n is the spectral index.

We assign `clusters' to the peaks of the background field according to the
following prescription:  each particle inside the cube is assigned a value
$\nu$ such
that $\delta_{\bf g} > \nu \sigma$, where $\sigma$ is the {\it rms} density
fluctuation within the cube and $\bf g$ is the nearest grid point to the
 particle. Here $\delta_{\bf g}$ is the density field smoothed with a
Gaussian filter function having a smoothing radius of 10 $h^{-1}$ Mpc.
The parameters $A$ and $n$ and $\nu$ must be chosen
such that the peak-peak correlation function in the range 10 -- 60
$h^{-1}$ Mpc, corresponds to the observed $\xi_{cc}$.
We find that for $A \approx 1.9 \times 10^{6}$, n$\approx -1.5$ and
$\nu\sim 1.3$ we obtain a distribution of simulated clusters with
$\langle r_{\circ} \rangle_{sim}=21$ $h^{-1}$ Mpc.
Furthermore these clusters have by construction non-zero higher
order correlations and we have found in PVJ, using the direct triplets-counting
method, $Q_{sim} \simeq 0.6 \pm 0.16$, where the uncertainty is the ensemble
one.

Finally we note that {\sl our simulated cluster catalogues have the same
geometrical boundaries, the same redshift and Galactic obscuration selection
functions, the same low-order correlation properties as well as the same mean
space density as the real clusters}.

\section{Method}
In order to obtain a continuous cluster density field we smooth the cluster
distribution in a $24^{3}$ cube ($480^{3}$
$h^{-3}$ Mpc$^{3}$) using a Gaussian kernel:
\begin{equation}\label{eq:ker}
{\cal W}(x_{i}-x_{\bf g})  = \left( 2 \pi R_{sm}^{2} \right)^{-3/2} \exp\left(
-\frac{|x_{i}-x_{\bf g}|^{2}}{2 R_{sm}^{2}} \right)
\end{equation}
The smoothed cluster density, at the grid-cell positions $x_{\bf g}$, is then:
\begin{equation}
\rho(x_{\bf g}) = \frac{\sum_{i} \rho(x_{i}) {\cal W}(x_{i}-x_{\bf g})}
{\int {\cal W}(|x-x_{\bf g}|) {\rm d}^{3}x}
\end{equation}
where the sum is over the distribution of clusters at positions $x_{i}$.
In order to study the cluster probability density function, $P(\rho)$, at
different
scales, we use four smoothing radii; $R_{sm} = 20$, 30, 40 \& 50 $h^{-1}$ Mpc
with $|x_{i}-x_{\bf g}| \le 3 R_{sm}$. In this case the integral at the
 denominator of eq.(8) is $\simeq 0.97$. There are, however, two problems which
we have to resolve:
\begin{itemize}
\item Due to the geometrical boundaries of the observation volume
and the zone of avoidance, each Gaussian sphere is {\em incomplete}.
Using a Monte-Carlo method, we estimate this incompleteness,
by defining a {\it grid completeness factor}:
\begin{equation}
f(x_{\bf g}) = \frac{\int n(x) p(x) {\cal W}(|x-x_{\bf g}|) {\rm d}^{3}x}
{\int n(x) {\cal W}(|x-x_{\bf g}|) {\rm d}^{3}x}
\end{equation}
where $n(x)$ is the density of a random distribution of points (thus $n(x)=
constant$) and  $p(x)$ represents the selection functions and geometric
boundaries. Note that by definition $0\le f(x_{\bf g}) \le 1$.
The smoothed cluster density is then corrected, at each grid point $x_{\bf
g}$, by $1/f(x_{\bf g})$.
However, to reduce the uncertainty due to the
approximate nature of this correction we use in our analysis only those
grid-cells with $f(x_{\bf g}) \ge 0.8$. Due to the small number of such cells
for large smoothing radii, we use also $f(x_{\bf g}) \ge 0.7$ but only for the
$R_{sm} \ge 40$ cases.
\item Discreteness effects can be introduced for small $R_{sm}$, if the
number of cluster counts, in the Gaussian sphere, is small or if the
smoothing fails to create a continuous density field. The latter occurs
only for $R_{sm} = 20$ $h^{-1}$ Mpc where some cells have $\rho_{sm}=0$.
This biases the $P(\rho)$ at $\rho/\langle \rho \rangle \ll1$. This brings
up the issue of which is the optimal smoothing radius to derive the
$P(\rho)$ of a discrete distribution of points. From considerations
presented in the following section we find that the optimal radius is of
the order the mean interparticle separation, ie. $\sim 30 - 40$ $h^{-1}$
Mpc for the Abell+ACO clusters.
\end{itemize}

By applying the selection functions and the geometric boundaries of the real
cluster distribution to that of the simulated clusters we have tested
the robustness of our correction procedure by comparing bin-by-bin the
original simulation cluster density contrast, $\delta_{ori}$ (whole cube),
with the recovered density contrast, $\delta_{rec}$.
We find an excellent agreement (Figure 1) and we have also
verified that the {\em pdf}'s,
before and after the application of the selection functions and the
correction procedure, are identical.

\section{Results \& Discussion}
In figure 2 we present the simulation-cluster mean {\em pdf}'s (over the
50 simulations), for all four
smoothing radii, after the application of the selection functions and the
subsequent corrections, discussed in the previous section. The errorbars are
the 1$\sigma$ scatter around the ensemble mean. We plot the Gaussian fit
(dashed line) to the cluster {\em pdf}, given by:
\begin{equation}
P(\varrho) = \frac{1}{\sqrt{2 \pi \sigma^{2}}}
\exp{\left[-\frac{(\varrho -1)^{2}}{2 \sigma^{2}} \right]}
\end{equation}
where $\sigma$ is the standard deviation of $\varrho (\equiv \rho/ \langle
\rho \rangle)$, and we also plot the log-normal fit, which has been found to
give an extremely good fit to the CDM density and the IRAS galaxy {\em pdf}
in the weakly-linear regime (Kofman et al. 1994). The log-normal distribution
is given by:
\begin{equation}
P(\varrho) = \frac{1}{\sqrt{2 \pi \sigma_{L}^{2}}}
\exp{\left[-\frac{(\ln \varrho - \mu_{L})^{2}}{2
\sigma_{L}^{2}} \right]} \frac{1}{\varrho}
\end{equation}
where $\mu_{L}$ and $\sigma_{L}$ is the mean and standard deviation of
$\ln \varrho$.

{}From figure 2 it is evident that the log-normal distribution fits well the
simulation cluster {\em pdf} for all four $R_{sm}$, while the Gaussian one
is rejected and although it becomes increasingly a better fit to the
cluster $P(\varrho)$, as $R_{sm}$ increases, it always provides a worse fit
than the corresponding log-normal distribution, even for $R_{sm}$= 50
$h^{-1}$ Mpc. Note, however, that for $R_{sm} = 20$ $h^{-1}$ Mpc and
$\varrho \lesssim 1$ the log-normal distribution does not fit well the
{\em pdf}. The reason is that at this smoothing scale and range
of $\varrho$ there are significant shot-noise contributions (see discussion
in previous section).

The log-normal distribution has been argued (Coles \& Jones 1991) to
describe the distribution of density perturbations resulting from Gaussian
initial conditions in the weakly non-linear regime. It has been shown to fit
well the Lick {\em
Euler-Poincar\'e} characteristic (a measure related to the genus) and the
2-dimensional Lick counts in cells at a
resolution which corresponds, at the characteristic depth of the Lick
catalogue, to scales of $\sim 4 - 10$ $h^{-1}$ Mpc (Coles \& Plionis 1991).
However, Bernardeau \& Kofman (1994) have shown that the log-normal
distribution is {\bf not} a universal form of the cosmic density {\em pdf} but
a very convenient fit only in some portion of the $(\sigma,$n)-plane
(ie, $\sigma \ll 1$ and n$\approx -1$). In this context it seems that
it is not a coincidence
that the simulation cluster {\em pdf} is fitted quite well by a log-normal
distribution but it could be a consequence of the fact that we used a power
spectrum with exponent near to $-1$ (n$_{sim} \approx -1.5$).
Alternatively, it could be a generic feature of high-peak biasing.

In figure 3 we present the one-point {\em pdf} of the Abell+ACO smoothed
cluster distribution, for three $R_{sm}$ values\footnote{We have not used
the $R_{sm} = 50$ $h^{-1}$ Mpc because there are only a few cells with
$f\ge 0.7$. In the simulation case we overcame this problem due to the
large number of available realizations.} and for two different values of
$f(x_{\bf g})$, the latter to check the effect of the limited number of
cells used to derive the {\em pdf}. The uncertainty here is given by
Poisson sampling errors and thus they should be considered only a lower
limit to the intrinsic uncertainty. A more reasonable indication of the
scatter is the ensemble errors seen in figure 2. Again we also
plot the best Gaussian (dashed line) and log-normal (solid line) fits. We
verify that also the real cluster smoothed {\em pdf} is fitted quite well
by a log-normal distribution, especially for $R_{sm} \ge 30$ $h^{-1}$ Mpc
where discreteness effects are minimized.

We now proceed to estimate the moments of the simulated and real cluster
distributions, which are defined by:
\begin{equation}
\langle \delta^{n} \rangle = \int_{-1}^{\infty} \delta^{n} f(\delta) d\delta
\end{equation}
where $\delta = \varrho-1$. Note that if $f(\delta)$ is a Gaussian then it is
defined in an infinite interval and thus it always has $f(\le -1) \neq 0$.
However by definition $\delta >-1$ and therefore $f(\delta)$
can be a Gaussian only in the limit $\sigma \rightarrow 0$. The gravitational
evolution of the $\delta$ field acts in a way to increase the variance,
$\sigma^{2} \equiv \langle \delta^{2} \rangle$, and thus in order for
$\delta > -1$ to
hold, $f(\delta)$ becomes skewed. Note that the skewness, $\gamma \equiv
\langle \delta^{3}\rangle$ of a Gaussian distribution (defined in an infinite
interval) is zero. In the case of the $\delta$ field, however, we have that:
\begin{equation}
\gamma = \int_{-1}^{\infty} \frac{\delta^{3}}{\sigma
\sqrt{2 \pi}} \exp{\left[-\frac{\delta^{2}}{2 \sigma^{2}} \right]} d\delta =
\frac{\sigma(1+2\sigma^{2})}{\sqrt{2\pi} \exp{(1/2\sigma^{2})}}
\end{equation}
and thus $\gamma \neq 0$ (unless $\sigma \rightarrow 0$).
We have already introduced the variance and the skewness of $f(\delta)$. In a
similar fashion we can define higher order moments; for example the 4$^{th}$
moment is the kurtosis, $K \equiv \langle \delta^{4} \rangle$ (cf. Lahav et
al. 1993; Kofman et al. 1994).

An important problem in determining the moments of $f(\delta)$ is the
contribution of discreteness effects which can dominate in low-density
discrete distributions. If the point-like distribution represents a Poisson
sampling of an underlying continuous density field then the shot-noise
contributions can be easily corrected (cf. Peebles 1980). However, the
galaxy-cluster distribution can be hardly considered a poisson sampling since
it rather represents a strongly biased sampling of an underlying galaxy
density field. Therefore
the poisson shot-noise corrections should not be expected to provide a
reasonable correction of the discreteness effects. In fact, Borgani et al.
(1994) found for their 2-d cluster analysis that indeed the poisson shot-noise
correction resulted in extremely noisy moments and they preferred not to
correct for shot-noise effects, which resulted in stable, well-behaved
moments.

Furthermore, the smoothing process itself suppresses the shot-noise
effects considerably (cf. Gazta\~naga \& Yokoyama 1993). From eq.(8) it is easy
 to show that the power spectrum of the smoothed cluster density
 is related to that of the cluster $P_{cl}(k)$ by
\begin{equation}
 P_s(k) = e^{-k^{2}R_{sm}^2} \left[ \frac {1}{N}+P_{cl}(k) \right] \simeq
 \frac {e^{-k^{2}R_{sm}^2}}{N} \left[1 + \frac {4\pi}{\langle l \rangle^3}
\int_{0}^{\pi/k} \xi(s)s^2 ds \right] \;\;,
\end{equation}
where $\langle l \rangle$ is the average cluster separation. Then the best
representation of $P_{cl}(k)$ is for $R_{sm} \simeq \langle l \rangle$, the
original field being oversmoothed for $R_{sm} \gg \langle l \rangle$ and
undersampled for $R_{sm} \ll \langle l \rangle$.

In Table 1 we present the second, third and fourth moment of the cluster
{\em pdf} as a function of smoothing scale, for the simulations as well as
the real data. It is apparent that the cluster moments are
significantly smaller than those of the simulation clusters. Although the
latter are constructed in such a way to reproduce the correct slope and
roughly the correct amplitude of the 2-point spatial correlation function of
the Abell and ACO ($R\ge 0$) clusters (see PVJ),
there are differences in the amplitude. For example the correlation length of
the Abell and ACO clusters is $r_{\circ} \simeq 18$ and 22 $h^{-1}$ Mpc,
respectively while of the simulated clusters $\langle r_{\circ}\rangle_{sim}
 \simeq 21$ $h^{-1}$ Mpc and since the Abell sample is the dominant one we
expect that
the overall cluster variance will be smaller than that of the simulated
clusters. In the same direction works also the fact that the simulated
clusters have a non-zero $\xi(r)$ at separations larger than those of
the real clusters and therefore since the smoothing process mixes a lot of
different scales, the non-zero $\xi(r)$ at large $r$ will tend to
increase the variance.

As we discussed in the Introduction many authors found, on the basis of a
variety of clustering scenarios from Gaussian initial conditions, that the
skewness is simply related to the variance of the mass distribution by:
\begin{equation}\label{eq:CF91}
\gamma = B \left(\sigma^{2} \right)^{A}
\end{equation}
This result holds for $\sigma \lesssim 1$ (cf. Fig.5 of Kofman et al.
1994) while $\sigma$ changes either
due to the clustering evolution or due to different smoothing radii. We
therefore plot in figure 4(a) the skewness, $\gamma$, versus the variance,
$\sigma^{2}$ for the real cluster {\em pdf} and in panel (b) for the
simulated cluster {\em pdf}. For the simulated cluster case we plot the
values for each of the 50 simulations and for each simulation at 4
different smoothing scales (as discussed in the previous section).
Similarly for the real clusters we plot the values corresponding to
different Abell-ACO normalization functions, $W(R)$, and for 3 smoothing
scales. It is evident that the plotted points in both, simulation and real
cluster, cases lie on the same line. To check whether eq.(~\ref{eq:CF91})
is fulfilled we have used a weighted least-square fit to determine the
slope and the intercept of the line:
\begin{equation}\label{eq:gamm}
\log \gamma = A \log\sigma^{2} + \log B
\end{equation}
and we have found for the cluster case:
$$A = 2.09 \pm 0.05 \;\;\;\;\;\;\; B = 1.94 \pm 0.15 \;,$$
and for the real cluster case:
$$A = 2.04 \pm 0.06 \;\;\;\;\;\;\; B = 1.8 \pm 0.2 \;,$$
which imply that indeed
eq.(~\ref{eq:CF91}) is fulfilled. This result is also in agreement with the
hierarchical relation ($\gamma = 3 Q \sigma^{4}$) with $Q \simeq 0.6$,
consistent with previous cluster correlation function analysis (cf. JV,
PVJ). The parameter $B$ is also called {\em normalized skewness}, $S_3$. In
general, the normalized moments are given by:
\begin{equation}\label{eq:cum}
S_{n} = \frac{\langle \delta^{n}\rangle_{c}}{\langle \delta^{2}
\rangle_{c}^{(n-1)}}
\end{equation}
where $\langle \delta^{n} \rangle_{c}$ are the cumulants. The normalized
skewness, $S_3$, plotted in Figure 5, for both the simulation and real
cluster {\em pdf} has roughly the same value, and is constant as a function
of $R_{sm}$. Such a behaviour is consistent with density distributions
evolving from Gaussian initial conditions (Coles \& Frenk 1991; Kofman et
al. 1994). Furthermore, eq.(~\ref{eq:CF91}) is valid also in the high-peak
biassing model in CDM universes (Coles \& Frenk 1991). This is worth
stressing since what we measure here is not the skewness induced on the
density field by gravity but that arising from a threshold peak-selection.

Therefore, the above results, together with the fact that both the simulated
and real cluster {\em pdf} are fitted by a log-normal distribution, imply:
\begin{itemize}
\item the Abell+ACO clusters are compatible with being high-peaks of a
Gaussian background and
\item the spectral index of density fluctuation spectrum, from which the
cluster distribution has emerged, is compatible with $\approx -1.5$.
\end{itemize}
We emphasize that they are {\em 'only'} consistent
because a thorough analysis of the expected behaviour of peaks of a
non-Gaussian background in 3-D has not yet been investigated thoroughly and
therefore we cannot exclude the possibility that such peaks may have a
similar behaviour.
However, there are indications from the study of the moments of simulated
{\em galaxy} distributions emerging from non-Gaussian initial conditions
(Weinberg \& Cole 1992; Coles et al. 1993), that they scale in a manner similar
to that of distributions emerging from Gaussian initial conditions. In fact,
Coles et al. (1993) find that in their initially positive skewed models,
$S_3$ is also a constant but has a higher value than that of the Gaussian
models. Initially negative skewed models are difficult to distinguish from
Gaussian models on the basis of their normalized moments since they produce
similar values of $S_3$ (see also Table 2 of Weinberg \& Cole 1992).

A similar analysis for clusters of galaxies has been performed, but only in
2-dimensions (Borgani et al. 1994) and it was found that not only the skewness
and the variance, in both Gaussian and non-Gaussian models, scale in a similar
manner but also $S_3 \approx 1.8$ for all the models. Since however the
projection from 3 to 2 dimensions Gaussianizes the distribution, as expected
from the central limit theorem, these results may not be representative of the
3-D case.

It would be interesting to compare the cluster {\em pdf} moments with those of
galaxies for similar smoothing scales. In fact, we have one smoothing scale
($R_{sm} = 20$ $h^{-1}$ Mpc) in common with the analysis by Saunders et al.
(1991) of the QDOT IRAS redshift sample, and it is this smoothing scale for
which they found large values of the variance and the skewness and claimed
that the CDM model is ruled out at a 97\% confidence limit.
Assuming that the linear biassing model is correct, on the
relevant scales, and using a biassing factor between clusters and IRAS
galaxies of $b_{cI} \approx 4.5$ (Peacock \& Dodds 1994) we predict
$$\sigma^{2}_{I,pred.} = b_{c,I}^{-2} \sigma^{2}_{c} = 0.0245 \pm 0.005$$
$$\gamma_{I,pred.} = b_{c,I}^{-3} \gamma_{c} = 0.0049 \pm 0.0018$$
while the corresponding measured values of the QDOT sample and of the CDM QDOT
look-alikes (from Saunders et al. 1991) are $(\sigma^{2},\gamma)$
$=(0.0669,0.025)$ and $(0.0192,-0.001)$ respectively. It is evident that the
predicted IRAS moments from the cluster
moments and from linear biassing are significantly smaller than the
Saunders et al. (1991) direct estimation but still larger, although
marginally so, than the CDM ones. Alternatively we can ask whether there is
a unique biasing value, $b_{cI}$, for which the rescaled cluster moments, to
the IRAS level, are equal to the QDOT-IRAS ones. In fact there is; using
$b_{cI}\approx 2.7$ we succeeded in the above task. However, such a low
value of $b_{cI}$ does not seem to be supported by many different studies
(cf. Jing \& Valdarnini 1993; Peacock \& Dodds 1994).

These results may be viewed as suggestive that the linear biassing model
cannot be applied straightforwardly to the moments of the {\em pdf}.
However, Juszkiewicz et al. (1993)
have found that eq.(~\ref{eq:CF91}) still holds for a biased population, if
the linear fluctuations are Gaussian and the biased density is a local
function of the matter density (see also Coles \& Frenk 1991).

Jing \& Valdarnini (1993) have found that the quadrupole anisotropy,
observed in the COBE experiment, can be reproduced by the measured Abell
cluster and IRAS galaxies power spectra, if their bias factors are $b_{c,I}
\simeq 3 - 4$ and $b_I \simeq 1.5$, respectively (see also Peacock and
Dodds 1994). These results, together with the value $S_3 \approx 1.8$ which we
obtain in this work for Abell+ACO clusters and using the linear biasing
framework, we estimate that for the matter distribution  $S_{3m} \simeq 8 -
12$.  These values look unrealistically large since Goroff et al. (1986)
 obtain $S_{3m} \simeq 3$ for a CDM spectra and a Gaussian smoothing
with $R_{sm} \ge 20$ Mpc.

Although, in the range of interest the cluster power spectrum has an effective
slope steeper than that of the CDM, the values of $S_{3m}$ cannot
exceed an upper limit of $\simeq 5$  obtained from perturbation theory
(cf. Fig.1 of Juszkiewicz, Bouchet \& Colombi 1993).
 Taken together all these results point toward a more cautious approach
for the linear bias assumption. On the other hand, however, the optical
cluster dipole (PV91; Scaramella et al. 1991) is found to be well aligned
with that of the CBR, which  means that clusters do effectively trace the
underlying gravitational field.

 We therefore suggest that the linear bias assumption is a good approximation
for spatial regions where $ \delta \gtrsim 0$ but breaks down in underdense
regions where $\delta \lesssim 0$. The skewness is a measure
of the asymmetry of the distribution with  respect the mean and therefore one
cannot simply relate the moments of the biased field to that of matter field
by a constant factor.

\section{Conclusions}

We have derived the Abell+ACO cluster probability density function at
different smoothing scales and we found it to be consistent with a
log-normal distribution at all scales considered. Similar results were
obtained from the analysis of synthetic clusters generated as high-peaks of
a Gaussian background density field having the same low-order correlation
properties and selection functions as the real clusters. Furthermore we
determined the moments of the {\em pdf} and we found them to scale
according to: $\gamma = S_3 \sigma^{4}$. Such a scaling is expected within
the framework of Gaussian initial conditions and gravitational instability.
However, since the clusters probe the linear field, the scaling is
attributed to the the fact that they are biased tracers of the underlying
density field. Furthermore, this scaling is in agreement with the
hierarchical clustering model with $S_3 \equiv 3 Q \approx 0.6$, consistent
with previous correlation function analysis (cf. VJ, PVJ).  These results,
together with the arguments of Kofman et al. (1994), indicate that the
observed cluster distribution is consistent with it being generated from a
initial Gaussian fluctuation spectrum.

\vspace{2cm}

\section*{Acknowledgements}
We thank Lev Kofman for a discussion, during his short visit at SISSA.

\newpage

\parindent 0pt
\section*{References}

Abell, G.O., 1958, ApJ, 3, 211

Abell, G.O., Corwin, H.G. \& Olowin, R.P. 1989, ApJS 70, 1

Bahcall, N.A. \& Soneira, R.M. 1983, ApJ 270, 20

Bardeen, J.M., Bond, J.R., Kaiser, N. \& Szalay, A.S. 1986, ApJ 304, 15.

Batuski, D.J., Bahcall, N.A., Olowin, R.P., Burns, J.O., 1989, ApJ 341, 599.

Batuski, D.J., Burns, J.O., Newberry, M.V., Hill, J.M., Deeg, H.J.,
Laubscher, B.E. \& Elston, R.J. 1991, AJ 101, 1983

Bernardeau F. \& Kofman, L.A., 1994, astro-ph/9403020

Borgani, S., Jing, Y.P. \& Plionis, M. 1992, ApJ, 395, 339

Borgani, S., Coles, P., Moscardini, L. \& Plionis, M., 1994, MNRAS, 266,
524

Bouchet, F.R., Scaffer, R. \& Davis, M., 1991, ApJ, 383, 19

Bouchet, F.R. \& Hernquist, L., 1992, ApJ, 400, 25

Bouchet, F.R., Strauss, M., Davis, M., Fisher, K.B., Yahil, A. \& Huchra, J
.P. 1993, ApJ, 417, 36

Cappi, A., Focardi, P., Gregorini, L. \& Vettolani, G. 1991, AAS, 88, 349

Coles, P. \& Frenk, C.S., 1991, MNRAS, 253, 727

Coles, P. \& Jones, B.J.T., 1991, MNRAS, 248, 1

Coles, P. \& Plionis, M. 1991, MNRAS, 250, 75

Coles, P., Moscardini, L., Lucchin, F., Matarrese, S. \& Messina, A., 1993,
MNRAS, 264, 572

Colombi, S., 1994, FERMILAB-Pub-94/050-A, submitted to ApJ Letters

Davis, M., Peebles, P.J.E, 1983, ApJ, 267, 465

Fairall, A.P. \& Jones, A. 1991, Publ.Univ. of Cape Town, No 11

Gazta\~naga, E. \& Yokoyama, J., 1993, ApJ, 403, 450

Goroff, M.H., Grinstein, B., Rey, S.--J. \& Wise, M.B. 1986, ApJ, 311, 6

Groth, E.J. \& Peebles, P.J.E. 1977, ApJ, 217, 385

Jing, Y.P.  \& Zhang 1989, ApJ, 342, 639

Jing, Y.P. \& Valdarnini, R. 1991, AA, 250, 1 $\;\;\;\;\;$ [JV]

Jing, Y.P. \& Valdarnini, R. 1993, ApJ, 406, 6

Jing, Y.P., Plionis, M \& Valdarnini, R. 1992,  ApJ, 389, 499

Juszkiewicz, R., Bouchet, F.R. \& Colombi, S., 1993, ApJ, 412, L9

Juszkiewicz, R., Weinberg, D. H., Amsterdamski, P., Chodorowski, M.
\& Bouchet, F. 1993; IASSNS-AST 93/50 preprint

Kaiser, N. 1984, ApJ, 284 , L9

Klypin, A.A., \& Kopilov, A.I., 1983, SvA Letters 9, 41.

Kofman, L., Bertschinger, E., Gelb, J.M., Nusser, A. \& Dekel, A., 1994,
ApJ, 420, 44

Lahav, O., Itoh, M., Inagaki, S. \& Suto, Y., 1993, ApJ, 402, 387

Mattig, W. 1958,  Astr. Nach. 284, 109

Muriel, H., Nicotra, M., Lambas, D.G. 1991, AJ, 101, 1997

Olive, K.A. 1990, Phys. Reports 190 , 307

Peacock, J.A. \& Dodds, 1994, MNRAS, 267, 1020

Peebles, P.J.E., 1980, {\em The Large Scale Structure of the
Universe}, Princeton University Press, Princeton.

Plionis, M. \& Valdarnini, R. 1991, MNRAS, 249, 46. [PV91]

Plionis, M., Valdarnini, R. \& Jing, Y.P. 1992, ApJ, 398, 12   [PVJ]

Postman, M., Spergel, D.N.,Satin, B., Juszkiewicz, R.  1989, ApJ 346 , 588.

Postman, M., Huchra, \& Geller, M. 1992,  ApJ, 384, 407

Raychaudhury, S. 1989, Nature, 342, 251

Rhee, G.F.R. \& Katgert, P. 1988, AAS 72, 243

Saunders, W. et al. 1991, Nature, 349, 32

Scaramella, R., Baiesi-Pillastrini, G., Chincarini, G., Vettolani, G. \&
Zamorani, G. 1989, Nature, 338, 562

Scaramella, R., Vettolani, G. \& Zamorani, G. 1991, ApJLett, 376, L1

Shapley, H. 1930, Harvard Obs. Bull., 874, 9

Struble, M.F. \& Rood, H.J. 1987, ApJS 63, 543

Sutherland, W. 1988,  MNRAS 234, 159.

T\'oth, G., Holl\'osi, J. \& Szalay, A.S. 1989,  ApJ 344, 65

 Vettolani, G., Cappi, A., Chincarini, G., Focardi, P, Garilli, B.,
Gregorini, L. \& Maccagni, D. 1989, AAS, 79, 147

Weinberg, D. H. \& Cole, S., 1992, MNRAS, 259, 652

\newpage

\begin{table}[h]
\centering
\caption[]{The low-order moments of the Abell+ACO (subfix $A$) and simulated
cluster (subfix $sim$) {\em pdf}'s, as a function of smoothing scale $R_{sm}$.
The uncertainty in
the real cluster moments is given from the scatter induced by using different
$W(R)$ functions (see text), while in the simulation case the
uncertainties are due to the scatter around the ensemble mean values. Note
that we also give the values for the fourth moment, $K$, although in the text
we do not discuss it due to the large uncertainty associated with its
estimation. Note, however, that for the $R_{sm} = 20$ $h^{-1}$ Mpc case,
where we have a relatively more reliable estimation, $S_4 (\equiv \frac{K - 3
\sigma^4}{\sigma^6}) \approx 5.1$ and 5.3 for the real and
simulated cluster data, respectively.}
\label{tab:mom}
\begin{tabular}{ccccccc}  \\ \\
$R_{sm}$ & $\sigma^{2}_{A}$ & $\gamma_{A}$ & $K_{A}$ &
$\sigma^{2}_{sim}$ & $\gamma_{sim}$ & $K_{sim}$ \\ \\

20 & 0.496$\pm 0.057$ & 0.4456$\pm 0.154$ & 1.363$\pm 0.7$ & 0.85$\pm 0.19$ &
1.44$\pm 0.68$ & 5.4$\pm 3.1$ \\ \\
30 & 0.136$\pm 0.023$ & 0.0327$\pm 0.0168$ & 0.365$\pm 0.566$ & 0.36$\pm 0.11$
& 0.29$\pm 0.2$ & 0.67$\pm 0.5$ \\ \\
40 & 0.048$\pm 0.013$ & 0.0040$\pm 0.005$ & $--$ & 0.2$\pm 0.078$ & 0.095$\pm
0.087$ & 0.173$\pm 0.154$ \\ \\
\end{tabular}
\end{table}

\newpage

\section*{Figure Captions}

{\bf Figure 1.}

Comparison, cell-by-cell, of the original simulated cluster fluctuation
field, $\delta_{ori}$, with the one recovered after applying the real
cluster selection functions and geometrical boundaries and using the method
described in section 3 (for $f(x_{\bf g}) \ge 0.8$). A random subsample of
10\%, from each of the 50 simulations, has been plotted.

{\bf Figure 2.}

The simulated cluster {\em pdf} at the 4 smoothing scales, indicated in each
panel (filled symbols). The errorbars are the scatter around the ensemble
mean (over 50 simulations). The solid line is the log-normal fit to the data
while the dashed line is the Gaussian fit.

{\bf Figure 3.}

The real cluster {\em pdf} at the 3 smoothing scales indicated in each
panel. Filled symbols correspond to $f(x_{\bf g}) \ge 0.8$ while open
symbols to  $f(x_{\bf g}) \ge 0.9$, for the $R_{sm} = 20$ and 30 $h^{-1}$
Mpc cases, respectively. For $R_{sm} = 40$ $h^{-1}$ Mpc they correspond to
0.7 and 0.6, respectively. The errorbars are Poisson errors and the line fits
are as in figure 2.

{\bf Figure 4.}

Variance versus skewness plot. The open circles, solid squares, open stars
and open squares correspond to $R_{sm} = 20$, 30, 40 and 50
$h^{-1}$ Mpc respectively. The solid line is the best fitted line given by
eq.(~\ref{eq:gamm}). (a) Real clusters (b) Simulated clusters.

{\bf Figure 5.}

The normalized skewness, $S_3$, versus smoothing scale for both real and
simulated clusters.

\end{document}